\documentstyle[preprint,aps]{revtex}

\preprint{Submitted to {\it Phys. Rev. E.}  See also http://mobydick.physics.utoronto.ca}

\begin{document}
\title{The Dynamics of Granular Segregation Patterns \\
in a Long Drum Mixer }
\author{Kiam Choo, Michael W. Baker, T. C. A. Molteno and Stephen W. Morris}
\address{Department of Physics, University of Toronto,
60 St. George St., Toronto, Ontario,
Canada, M5S 1A7}
\date{\today}
\maketitle

\begin{abstract}

We have studied the early time evolution of granular segregation patterns in a horizontal rotating cylinder partially filled with a sand/salt mixture.  The growth of concentration fluctuations starting from premixed initial conditions was analyzed using Fourier techniques.  In one mixture, we observed generally merging dynamics in the segregated bands up to the onset of saturation.  At the threshold of saturation, we found a spectrum of wavelengths with a broad peak.  The peak position was nearly independent of the tube rotation rate.  The overall growth rate of Fourier modes had a maximum at a particular value of the angular rotation frequency.  In a slightly different mixture, we observed transient traveling waves when the larger grains were in the majority.  We measured the wave speed as a function of several parameters using presegregated initial conditions to launch waves of various wavelengths.

\end{abstract}

\section{Introduction} 
 
Heterogeneous mixtures of granular materials tend to segregate by size or shape under conditions where one might naively expect mixing. In 1939, Oyama noticed that a binary mixture of large- and small-grained sand segregated into axial bands when tumbled in a horizontal ``drum mixer''.\cite{oyama}  After a hiatus of several decades, this counterintuitive phenomenon has captured the attention of many researchers,\cite{williams,bridge,donald,dasgupta,savage,nakagawa,nakagawa-mri,zik,hillpre,mri,hillevolution,chicharro,ourPRL,frette} and it has become something of a standard problem in the study of granular materials.  
 
The experiment is deceptively simple.  In a typical arrangement, equal amounts of two grain sizes, often of contrasting colors, are uniformly mixed and used to partly fill a horizontal transparent tube.  The tube is then rotated about its long axis at a rotation rate which is sufficient to cause the grains to stream smoothly.  Axial bands of segregation first appear after a few tens of rotations and are faint at first, but intensify and saturate. After a few hundred to a few thousand rotations, an apparently stable array of axial bands of roughly equal spacing is formed.  Bands may merge or split on long time scales, but the general tendency is towards merging\cite{dasgupta,nakagawa,chicharro,frette}.  With very extended rotation time, some experiments show that the bands exhibit a remarkably long period of metastability, but eventually all merge and finally achieve complete segregation, where the two species occupy opposite ends of the tube.\cite{chicharro} 
 
The occurrence of axial segregation seems to require that the dynamic angle of repose (roughly, the angle of the flowing surface) be different for the two species \cite{dasgupta,savage,hillpre}.  This has been taken to be evidence that the geometry of surface flows is responsible for the segregation.   This observation has inspired theoretical models\cite{savage,zik} that treat the onset of segregation as a kind of reverse-diffusion driven instability in which initial concentration fluctuations are amplified.  
 
Many researchers have studied the number of bands that appear.  It is claimed that more bands appear when more small particles are used \cite{dasgupta}, and that more bands appear when the rotation rate is slower\cite{nakagawa}.  Recently, MRI studies have shown that there exist subsurface bands that may not be visible from the surface\cite{mri,hillevolution}.  This suggests that visual band counting may be missing some bands, and casts doubt on a purely surface-driven segregation mechanism.  In some systems, the segregation also appears to be affected by coherent surface avalanche waves\cite{frette} or fingering instabilities\cite{thoroddsen}.  These would provide non-diffusive mechanisms for axial segregation.
  
In this paper, we report our observations on the dynamics of segregation in a mixture of sand and salt in a long pyrex tube. We have focussed on the very early stages of the evolution, in an effort to shed light on the nature of the instability that leads to segregation.  The primary improvements we made over previous experiments were in control of the initial conditions and improved resolution in the image processing and Fourier analysis. By employing a very long tube, a larger number of bands can be studied, yielding Fourier peaks that are well separated from dc and have good resolution.  We digitized images of the flowing surface and used averaging and image-division techniques to observe faint bands.   

We studied the initial growth of the bands by observing the exponential growth of Fourier modes.  At the threshold of saturation, we examined the spectrum of preferred wavelengths of the band pattern.  We measured the rate at which the bands grew and identified the dominant wavelength, as a function of the rotation rate of the tube. 

By composing images of the spacetime evolution of the surface contrast, we were able to discern transient traveling waves\cite{ourPRL} and other structures.  In this paper, we characterize the traveling wave speed as a function of wavelength, rotation period, filled fraction of the tube, and the relative proportions of the two granular components.  These traveling waves are surprizing because they are not expected from any simple diffusion model of segregation.

This paper is organized as follows. In section \ref{expt}, we describe the apparatus and experimental procedures.  In section \ref{results}, we describe the detailed observations.  Section \ref{discussion} and \ref{summary} contain some discussion and a brief summary of our conclusions.

\section{Experiment} 
\label{expt}

In this section we describe the apparatus, the experimental procedures and the data acquisition and analysis.  The experiment was optimized for the quantitative observation of the early stages of segregation.  Many parameters effect the details of the phenomena, and we were only able to systematically explore a few of them.

\subsection{the tube}
 
The drum mixer consisted of a $\sim$1 m long tube partially filled with the granular mixture which was rotated about its long axis at a constant angular frequency $\omega$ using a computer-controlled stepper motor. 
The tube was made as long and narrow as practical so that the number of bands was large.  This maximized the resolution of Fourier analysis, described below.  Three different tubes were used.  Each had inner diameter 27 mm.  Most of the observations were made using one of two pyrex tubes which had aspect ratios (length/diameter) of 40.6$\pm$0.2 and 38.6$\pm$0.2.  The third tube was made of plastic, and was only used to investigate charging effects.   Friction with the endcaps tends to produce segregation; we used teflon-covered end plugs in an attempt to minimize this effect. 

The speed of the traveling waves we observed with some mixtures was sensitive to the precise leveling of the tube. Waves traveled downhill more quickly than uphill.  We leveled the tube to better than $0.1^\circ$ by matching the speeds of the left- and right-going waves.

\subsection{the granular materials}

The components of the granular material were selected to maximize the color contrast between the two species.  By doing this, we were able to observe very faint bands.  This capability was especially important for studying the traveling waves.  The characteristics of the grains were by far the most difficult parameters to specify and control.

The smaller black component was ``hobby'' sand, while the larger white component was common table salt.\cite{thegrains} Under magnification, we found that most of the salt grains had a cubic crystalline shape and only a very small fraction were irregularly shaped.  The black sand grains, on the other hand, were all somewhat rounded and were very irregularly shaped. We also performed some tests with light colored Ottawa test sand in place of the salt.  This sand was very rounded in shape.
 
Two different types of mixtures were used for the experiments, which differed in the internal size distributions of each component.  These distributions are shown in Figure \ref{sandsalt-distribution}, and were determined using standard sieves.  In all cases, the sand was smaller than the salt.  Mixture B had a larger variety of particle sizes, and had larger salt grains on average than mixture A.  These sizes were selected to be roughly similar to those used by Das Gupta {\it et al}. \cite{dasgupta}

We define the size composition fraction $\phi$ to be the volume of the salt component divided by the sum of the volumes of the sand and salt components, where all volumes were measured {\it before} mixing.  Because of packing effects, the volumes of mixtures were typically about 10\% smaller than the sum of the volumes of the unmixed components.  Thus, $\phi$ is not exactly equivalent to the volume concentration of the salt in any mixture, but is experimentally easier to specify and vary systematically.  In the runs described below, mixture A was used at a fixed $\phi = 0.50$, while type B's mixture ratio was varied between 0.33 and 0.79.

\subsection{filling the tube}

Filling the long tube with the mixture to the desired amount was a nontrivial task.  Any sort of pouring, flowing or shaking operation causes segregation.  We also wanted to avoid nonuniformities in the initial filled volume fraction along the length of the tube.
  
The tube was filled by first gently placing the sand/salt mixture in a U-shaped aluminum channel as long as the tube, inserting it lengthwise into the tube, and then rotating it to dump its contents in place.  In most runs, two such operations were used, corresponding to a filled volume fraction $\eta$ of 0.28 $\pm$ 0.02.  This quantity is necessarily somewhat ill-defined, as packing effects, dilation under shear and segregation all change the volume of the granular mixture during the course of the experiment.  In some runs, we varied the filled volume fraction by filling the tube with different numbers and sizes of U-shaped channels.  In these cases, we simply used the total mass of material in the tube as a measure of the filled volume fraction.
 
Mixture A with $\phi = 0.50$ and $\eta = 0.28$ was used with randomly premixed initial conditions for studies of the dynamics of segregation into normal bands.  The initial conditions were prepared by manually mixing the grains in small quantities and loading the tube as described above.  In spite of our best efforts to mix the ingredients, there were typically still some concentration fluctuations that were large enough to be visible to the eye. 

Mixture B with a range of $\phi$ and $\eta$ was used for studies of the dynamics of the transient traveling waves.  For this purpose, presegregated initial conditions were prepared for a series of runs.  These were obtained by placing thin partitions in the U-shaped channel and then alternately filling each segment with one component of the mixture.  When the channel was full, the partitions were removed and the tube was loaded as before.  By varying the position of the partitions, we could prepare any desired wavelength of initial segregation, while holding the size composition fraction $\phi$ nearly constant. To determine the exact value of $\phi$, the mixture was sieved after each presegregated run.  We used 100\% presegregated bands as initial conditions for our study, but any other degree of presegregation could also be made by filling the segments of the channel with suitably premixed material.

\subsection{electrostatic and grinding effects}

Given the highly insulating nature of all the components, electrostatic effects must be carefully considered. Charge transferred between the grains or onto the tube can modify the segregation patterns by introducing long-range forces.\cite{electro}  Rubbing the exterior of the rotating tube or repeated rapid pouring of the mixture when it was outside the tube caused a few of the grains to stick to glass surfaces. We monitored the humidity of the enclosure surrounding the experiment with a hygrometer, and it was found to vary from 18 - 34\%.  No sticking due to static charging or humidity-dependent effects were observed in normal runs, which lasted $\sim$ 1 hr. In very long runs, which we do not discuss here, static charge did eventually build up enough that sticking became apparent. To investigate the effect of charge on the tube, we made a few runs using a plastic tube in place of the pyrex one.  With the plastic tube, charging effects were much greater, but all of the same phenomena, including the traveling waves, were observed.  This indicates that although charging effects may be present at some level in the pyrex tube, they do not significantly effect our results.
  
The grinding effect of repeated tumbling caused slow but systematic changes in the quantitative behavior of the mixtures, particularly of traveling waves. The salt became progressively greyer in color as the polishing of the grain's surfaces made them more transparent.  There was also a tendency for the corners and edges of the salt grains to become damaged, making the grains slightly rounded.  There was no obvious change in the sand component with grinding.  In no case was grinding sufficient to cause any measurable change in the size distributions, as determined by sieving. To avoid systematic effects, we used mixtures that were ``run-in'' briefly to avoid initial transients and then discarded them after $\sim 10$ hours of tumbling.

\subsection{image acquisition and analysis}
 
During rotation, the surface brightness of the flowing material was visualized using a monochrome CCD camera that could image the entire length of the tube using a wide angle lens (focal length 3.7 mm). The tube was uniformly lit by a single long fluorescent tube, which was located just above it. The video signal was acquired by a computer via an 8-bit frame grabber.  The camera was positioned so that it was pointed perpendicular to the flowing surface.  Only the portion of the video frame covering the flowing surface was digitized.  After acquisition, the resulting long rectangular image was corrected in software for the geometrical distortion introduced by the wide angle lens so that length measurements could be accurately made. In this step, we also interpolated the 400 pixel length of the array to 512 pixels, to facilitate later Fourier analysis. Spatial intensity variations introduced by nonuniform lighting were removed by dividing by a reference image.  Each rectangular image was then spatially averaged across the height of the streaming surface to produce a one-dimensional greyscale array. In this step, we also interpolated the 400 pixel length of the array to 512 pixels, to facilitate later Fourier analysis. Arrays taken at regular time intervals were stacked to form spacetime images which recorded an entire run. Calibration of the greyscale with respect to the local volume concentration of black sand indicates that, except for low black concentrations ($<$  20\%), the greyscale is, to a good approximation, linearly related to the local salt concentration. 

The spacetime images were analyzed with one- and two-dimensional Fourier transforms.  A one-dimensional Fourier transform, applied separately to each row (i.e., in the space direction only), was used to study the time evolution of the Fourier modes over the course of a run.  Two-dimensional Fourier transforms of the whole spacetime image were used to analyze traveling wave states to obtain wave speeds and for simple image processing to extract left- and right-going waves.

\section{Results} 
\label{results}
 
We observed radial segregation, end segregation, axial segregation and transient traveling waves, under various experimental conditions.  The latter two were studied in detail and are described in the following sections.  Except for traveling waves\cite{ourPRL}, all of these effects have previously been observed in other experiments. \cite{bridge,donald,dasgupta,savage,nakagawa,zik,hillpre,hillevolution}

 For our system, radial segregation occurs rapidly, on a time scale of $\sim$ 10 rotations.  We cannot observe this directly, but it is evident from a change in the overall surface color as a portion of the smaller black component migrates to the axis of rotation. As in many previous studies, we also observed the formation of end segregated bands that were also composed of the smaller black component. These  formed in spite of our teflon-covered end plugs.  However, if a white salt band was started against an end cap in a presegregated run, then the band stayed white.  The end segregated bands formed more quickly than other bands, but did not subsequently change over the duration of the run. 
 
\subsection{the formation of normal bands} 
 
In order to study the growth and saturation of normal bands, we used mixture A at a fixed composition $\phi = 0.50$ and filled volume fraction $\eta = 0.28$.  The angular rotation frequency $\omega$ was varied between 1.57 and 8.38 rad/s. This interval covers essentially the entire range over which the mixture streams smoothly down the flowing surface.  At lower $\omega$, intermittent avalanching is observed, while for larger $\omega$, the mixture spends part of each rotation airborne.  A run typically lasted for 600 seconds, with 5 seconds elapsing between image acquisitions. We used only premixed initial conditions. A spacetime image of a sample run is shown in Figure \ref{sample-dynamics}.  Several images are averaged together to form each line of the picture.

The first faint axial segregation bands can be seen after just a few tens of rotations.  These bands usually intensified until the black bands saturated at 100\% concentration.  Since earlier radial segregation carries smaller black sand to the axis of rotation, it seems likely that the saturated black bands seen on the surface consist only of that species all the way through.  On the other hand, the apparently saturated white bands probably still contain a black core that is not visible from the surface. In the following, we will focus on the dynamics of the growth of the black bands.

At the highest rotation rates, $\omega > 7 $rad/s, the black bands apparently saturate but do not reach 100\% concentration.  The bands stayed faint.  Apparently the randomizing diffusive effects of high rotation rates limit the final concentration of the bands.  On the other extreme, at very low rotation rates, $\omega < 1.6 $ rad/s, the time to saturation sometimes exceeded our observation time. Bands do not all saturate at the same time. If an area has a high local black concentration at the start of a run, saturation is reached earlier there. 
 
Saturated black bands are metastable over several thousand rotations.  They were never observed to split.  Only a few merging events were observed between saturated bands, and they occurred very slowly. Saturated black bands can sometimes be seen to absorb unsaturated ones.  The traveling waves discussed in the next section also tend to be absorbed by saturated bands.
 
Before saturation, bands exhibit more interesting dynamics. Bands can appear and disappear, and they can merge and split.  Examples are shown in Figure \ref{sample-dynamics}. 
Fourier spectra with multiple modes with no one mode dominant are common.  The growth of the total power in the Fourier spectrum is well-described as being initially exponential, as illustrated in Figure \ref{log-totalpow-with-spacetime}.  We omit the power in the first 5 Fourier bins to eliminate the dc component.   The total power ceased to grow exponentially at the same time as saturated bands appeared. Figure \ref{growthrate-vs-freq} shows the exponential growth rate of the total power {\it vs.} $\omega$.  We express the growth rate in units of the rotation period in order to eliminate the obvious tendency for growth to be faster for larger $\omega$. Interestingly, the growth per tube rotation appears to reach a maximum at $\omega \simeq 4 $ rad/s. 
   
To look for characteristic spacings of the saturated bands, the Fourier spectra were examined at the onset of saturation, for various $\omega$.  The onset of saturation was determined by the loss of exponentiality of the total power.  Examining the Fourier spectra before saturation is not particularly informative as the relative strengths of Fourier modes change rapidly due to band dynamics. 
The spectra at saturation are noisy; to get some insight, we simply averaged together spectra for all $\omega$ in the range 1.57 - 8.38 rad/s.  Each spectrum was normalized by dividing by the total power outside dc before being included in the average.  Figure \ref{average-fft} shows the result. There is a clear peak near a band spacing of 50 mm.  There is a rather broad range of growing modes about this peak, however, with the hint of other peaks.

For each $\omega$, we also examined the spectra to ascertain which Fourier mode, if any, dominated.  As there were sometimes multiple peaks of almost equal size, centre-of-mass wavelengths were also obtained to more accurately take the spread into account.  The centre-of-mass was obtained over a range of wavenumbers from just above dc to twice the wavenumber of the highest peak.  The result is shown in Figure \ref{selected-wavelength}.  It shows that, to a first approximation, the selected wavelength of the axial bands is independent of $\omega$.

\subsection{transient traveling waves} 
 
Under certain conditions, mixture B exhibited traveling waves during the transient that precedes the formation of the usual stationary axial bands.  Here we review this puzzling phenomenon, which was first described in Ref. \cite{ourPRL}, and make some further observations. 

Traveling wave transients occur only in a restricted region of the parameter space in which axial segregation is found.  Outside this region, one finds generally merging transients like those shown in Fig. \ref{sample-dynamics}. Our objective was to determine, as far as possible, which parameters are important for the existence of wave-like transients and which are not.  Surprisingly, the size composition fraction $\phi$ turns out to be an important parameter.  We varied $\phi$ over the range 0.33 - 0.79 in mixture B, while $\omega$ and $\eta$ were held constant. The waves were not observed unless $\phi > 0.55$, and, for a given wavelength, their speed increased with $\phi$.\cite{ourPRL} Large $\phi$ corresponds to mixtures rich in the larger salt component. The traveling waves obey a well-defined dispersion relation.  On the other hand, we show below that the traveling waves are rather insensitive to variations in $\omega$ or $\eta$.   

With premixed initial conditions and $\phi \sim 0.6$, patches of spontaneous left- and right- traveling waves are observed, which eventually give way to the usual stationary bands. The waves have a preferred wavelength of 45 $\pm$ 5 mm.  The waves move at nearly constant velocities in either direction and usually pass through each other with little or no interaction, until band formation, which has been occurring in parallel, grows strong enough to interfere with and destroy them. The traveling waves are dissipated in the vicinity of black bands which are saturated at 100\% sand.  All these features are illustrated in Fig. 1(a) of Ref. \cite{ourPRL}.

In order to study homogeneous regions of the traveling waves, we employed presegregated initial conditions in which a chosen wavelength was launched all along the length of the tube.  The resulting uniform traveling waves pass through each other to form a standing wave which persists for several oscillations before it is disrupted by stationary band formation. Figure \ref{seeded} shows a spacetime image of a portion of such a standing wave. 

In order to study how the traveling velocity varies with wavelength, numerous runs were performed with $\phi=0.67$, $\eta = 0.28$, $\omega = 4.841$ rad/s and different presegregated initial wavelengths. The velocities of the resulting waves were measured by locating the times when the standing waves pass through zero. To avoid early transients and later nonlinear effects, we used only the positions of the first few zero-crossing times.\cite{ourPRL}  We also measured the speeds of some spontaneous waves occurring with premixed initial conditions at these values of $\omega$, $\phi$ and $\eta$ to confirm that they were similar to the presegregated waves.  The speeds of spontaneous waves were estimated from the slopes of their worldlines in spacetime images, and by extracting the positions of peaks in 2D Fourier transforms of spacetime images.  Figure \ref{dispersion-relation} shows the dispersion relation obtained from velocity and wavelength measurements.

An interesting feature of the dispersion relation is that the waves do not travel when presegregated below a certain cutoff wavenumber.  We refer to these as ``frozen'' bands. The following observations indicate a possible explanation for this behavior.  When the rotation is started, there is a brief relaxation phase during which the initially sharp presegregated dark bands become diffuse and narrower as the material near the black/white interfaces mixes.  It seems that traveling waves occur only when this mixing is sufficient to completely submerge the initial presegregated black bands.  For wavelengths which do not travel, it was found that the central portion of the black bands remained 100\% sand. Salt cannot move through such pure regions of sand, and thus broad sand bands cannot travel.  The appearance, immobility and stability of these frozen black bands suggests that they are similar to the 100\% segregated bands that eventually grow up spontaneously and halt the traveling wave transient regime.
 
The transition region between frozen bands and traveling waves is difficult to pin down experimentally. In this vicinity we find that some bands remain frozen, while others travel a short distance, but merge when they encounter another band (Figure \ref{merging}).  These mergings appear to be similar to the late-stage absorption of traveling waves by 100\% segregated bands, seen at shorter wavelengths. 
  
Using the greyscale as a measure of the local concentration, we can examine the amplitude linearity of the traveling waves.  Their amplitude is quickly damped away with rotation, as shown in Fig. \ref{vamplitude}(a).  Despite this large change in amplitude, their velocity is constant, as shown in Fig. \ref{vamplitude}(b).  Thus, the waves are remarkably linear.  This is also indicated by the fact that waves traveling in opposite directions pass through each other with little or no distortion, i.e., they obey superposition.  The merging behavior seen in Fig. \ref{merging} can be interpreted as due to a loss of linearity near the transition from traveling waves to frozen bands. 
  
The existence and velocity of the traveling waves are robust over a wide range of filled volume fractions $\eta$, as shown in Fig. \ref{speed-vs-filling}.  The wave speed is essentially constant, even when $\eta$ is changed by a factor of two.  The accessible range of $\eta$ is limited by the tendency of the material to slip along the walls of the tube for small $\eta$, and the need to leave enough free volume to fill the tube using the U-shaped channel at large $\eta$. 

We have also measured the traveling wave speeds as a function of the angular frequency of rotation $\omega$. We found that bands in a presegregated run freeze for $\omega > 5.464$ rad/s, but travel for smaller $\omega$. Fig.  \ref{speed-vs-period} shows the wave speed vs. the period of rotation $T=2 \pi/\omega$.  When velocity is expressed in terms of distance traveled per rotation period, we find that wave speeds do not have a strong dependence on $T$.

To check for momentum effects, we stopped the rotation of the tube during the propagation of a standing wave, and then restarted it when all grain motion had ceased. This did not significantly alter the behavior of the waves, which continued to travel once the rotation restarted.

We examined how the traveling wave phenomenon depends on the details of the two size distributions and on grain shape.  In general, almost any change in either of these quantities causes the traveling waves to change or disappear altogether.  Substitution of the cubical salt component with the same size distribution of rounded Ottawa sand resulted in segregation but had transients with no traveling waves. Similarly, even the rather minor rounding effect that prolonged grinding has on the salt grains eventually results in the disappearance of the traveling waves.  Variations of the internal size distribution of either of the components of mixture B has a similar effect. These observations suggest that traveling waves occupy only a small portion of the parameter space over which segregation itself exists.

\subsection{other traveling structures} 
 
Some other interesting dynamical objects can be observed in mixture B. Fig. \ref{fountain} shows a  ``fountain'' band which sent out traveling waves in both directions. The fountain appears to constantly send out traveling waves that disappear as they near the 100\% segregated bands on either side.  If the 100\% segregated bands indeed absorb traveling waves, then it is possible that, eventually, the fountain would become depleted and cease.  However, that did not happen during the $\sim 2000$ rotations that we followed the evolution of the fountain.  The fountain occurred spontaneously in a presegregated run; it is not known how to reproducibly make one.

Using presegregated initial conditions, it is straightforward to launch pulses.  Fig. \ref{pulse} shows the evolution of one such pulse, which was prepared by laying down three bands, two white and one black, of 100\% presegregated material in an otherwise premixed background with $\phi=0.67$.  When the rotation was begun, each of the two prepared white bands pulled in black sand on their outer boundaries, and proceeded to travel away from each other.  These two waves appear to be accompanied by fainter, co-traveling waves on their sides.  It would be interesting to systematically study various kinds of pulses and their interactions.    

\section{Discussion} 
\label{discussion}

In the absence of a general theory of granular matter, any explanation of axial segregation must be developed heuristically.  Since the grains are more-or-less locked in solid rotation for most of the rotation period, models have naturally focussed on identifying sorting processes occurring in the dilated, flowing surface layer.  In the simplest picture, the surface is characterized by a single number, the dynamic angle of repose $\theta$. It is readily observed in most segregating mixtures that axial segregation results in a modulation of the surface slope along the axis of the tube.\cite{donald,dasgupta,hillpre,zik}  Thus, one assumes that $\theta$ depends on both $\omega$ and the local concentration $c$, and seeks a dynamical equation for $c$.\cite{savage}  In one dimension, the conservation of sand implies the continuity equation 
\begin{equation} 
\partial_t c = -\partial_z J, 
\label{continuity} 
\end{equation}  
where $z$ is the axial coordinate of the tube and $J(c,z,t)$ is the axial concentration current.  The current $J$ is assumed to have the form 
\begin{equation} 
J = (\beta - D) \partial_z c, 
\label{current} 
\end{equation} 
where $D>0$ is the usual Fick diffusion coefficient and $\beta(c,x)$ depends on the difference in the dynamic angles of repose of the two components. If $\beta > D$, Eqn. \ref{continuity} is a diffusion equation with a negative diffusion coefficient, so that initial concentration fluctuations grow.

Several observations support this general approach.  Das Gupta {\it et al.} \cite{dasgupta} and Hill {\it et al.} \cite{hillpre} have observed that segregation stops, and mixing occurs, when $\omega$ is such that the difference in $\theta$ between the two species is zero.  Segregation is also observed in two-dimensional poured piles\cite{makse}, where it has been explained by various models based on differing angles of repose.\cite{boutreux,stanley}  

An elaborate theory of axial segregation based on angle of repose effects has been proposed by Zik {\it et al.} \cite{zik} in which an expression for $\beta(c) \propto c(1-c)$ was derived from geometrical considerations and continuity.  Although this model has its successes, such as obtaining an S-shaped cross-sectional profile of the flowing sand that is qualitatively similar to experimental observation, it can do no more than indicate that an instability occurs.  Without nonlinear terms, it cannot make any prediction about the saturation or preferred wavelength of the band pattern. However, if we assume that saturation effects with a short-wavelength cut-off eventually limit the growth of the fluctuations, it follows that axial segregation proceeds in a manner analogous to spinoidal decomposition.  The initial stages exhibit an exponential growth phase, followed by a slow merging of nearly 100\% segregated bands.  This is qualitatively what we observe.

Our data suggest that there exists a fastest growing fluctuation, and it has a wavelength of about 50mm, or about twice the tube diameter.  This wavelength is essentially independent of $\omega$, while its growth rate is a maximum for $\omega \sim 4$rad/s.

The irreversible evolution of the bands favours merging and will result in complete segregation after a sufficiently long time.  Such behavior has been observed in some mixtures: after two weeks of rotation, Chicharro {\it et al.} \cite{chicharro} reached a fully segregated configuration in which the two species each occupied one end of the tube.  They also observed that the almost regular initial spacing of the bands was surprisingly metastable, however.  Other experiments\cite{frette} have clearly observed late-stage dynamics in which segregation is not complete.

In the negative-diffusion model, segregation is accomplished by diffusive differential surface transport down axial gradients of the angle of repose.  It is clear from various experiments that this picture is highly simplistic, however.  It ignores the radially segregated core of the smaller component which lies below the surface.  Recent MRI experiments\cite{mri,hillevolution} have imaged the core, and shown that subsurface segregated bands exist, which presumably cannot be due to purely surface sorting.  Transport to and from the core can be driven by the same mechanism which drives the fast radial segregation.  This mechanism has been elucidated in a series of experiments\cite{2dexpt} and simulations\cite{2dtheory} on the simpler case of two-dimensional drums partly filled with disks.  While it is not at all obvious how material in the core might move axially, it is clear that the core represents a reservoir of material which is coupled to the surface via a channel which is not taken into account in the negative-diffusion model.

The traveling waves we observed in mixture B when $\phi$ is sufficiently large are also irreconcilable with the negative-diffusion model. It is not possible to support bidirectional, linear traveling waves in a one dimensional pde which is first order in time.  No such equation can describe counter-propagating waves which together form a standing wave, as we observe.  This is most easily seen by considering the moment when the standing wave passes through zero as an initial condition for the rest of its evolution.  Obviously, the subsequent motion must depend on a momentum-like quantity which is specified at that instant as well as on the zero displacement initial condition. Since equations which are first order in time lack any analog of momentum initial conditions, they cannot be sufficient.  Such waves are not, however, inconsistent with first order dynamics in more dimensions, or, for example, with coupled first order equations.  Such equations might arise in models in which the core is taken into account, or which involved a more realistic two-dimensional surface.  

Collective motions of the grains provide another mechanism which might affect transport in a non-diffusive way and cause momentum-like terms in the concentration dynamics. In experiments with homogeneous sand\cite{fauve,capon}, oscillatory cellular instabilities of the streaming surface and undulations modes of the lower contact have been observed. These occurred on a timescale comparable to one rotation period and are associated with waves of avalanching propagating axially along the flowing surface. The time-scale of our traveling concentration waves we observed is at least 100$\times$ longer.  In other experiments\cite{frette}, long sequences of avalanche waves are clearly correlated to segregation. The avalanche waves can have a strong effect on slow axial segregation phenomena in mixtures where one component exhibits avalanche waves and the other does not. This is true despite the very large separation of timescales involved.  So far as we can determine, none of these avalanche waves are present in our experiment, although they are conceivably present at some amplitude which is too small to detect visually.  Stopping and starting the tube rotation did not prevent the continued propagation of the traveling waves, but this only interrupts the instantaneous momentum of the grains and does not necessarily mean that avalanche waves are not involved.    Very thin layers of sand, corresponding to filled volume fractions $\eta \approx 0.01$ have been discovered\cite{thoroddsen} to have a rich dynamical behavior which is reminiscent of that of fluids in a similar configuration\cite{valette}.  Various collective modes are observed, including traveling waves.  Thus, it seems possible that non-diffusive transport due to subtle collective motions of the grain may couple to the concentration field and give rise to some of the traveling wave effects we observe. This possibility deserves further investigation.
 
\section{Summary and Conclusion} 
\label{summary}

We studied the initial growth and saturation of patterns of axial segregation of a binary granular mixture in a rotating tube.  By Fourier techniques, we followed the mode structure of the pattern as it emerged from random, premixed initial conditions.  We measured the growth rates of various modes and determined that the system experiences initially exponential growth across a broad spectrum of modes peaked at a wavelength of 50mm.  This preferred wavelength was largely independent of rotation rate.  We have also found that the overall growth rate of structures varies with rotation rate, with a maximum at $\omega \sim $4 rad/s.

In addition to normal axial segregation, we have further examined the phenomenon of transient traveling segregation waves that occurred in certain mixtures. We have measured the velocity of the traveling waves and its dependence on the size composition fraction, the wavelength, the rotational frequency and the filled volume fraction.  We have found transitions from traveling to stationary bands as a function of some of these parameters.  Between the traveling regime and the stationary regime, we find merging behavior. We established the amplitude linearity of the waves. We have also observed traveling pulses and sources.  

The rich phenomenology of axial segregation presents many opportunities for quantitatively testing models of granular segregation. 

\acknowledgements

We would like to thank Elaine Lau, Eamon McKernan and Holly Cummins for development work, and Zahir Daya, Wayne Tokaruk and Troy Shinbrot for useful discussions.  This work was supported by The Natural Sciences and Engineering Research Council of Canada.

\begin{figure} 
	\caption{Histograms of sand and salt sizes in mixtures A and B. Note the different scales.} 
	\label{sandsalt-distribution} 
\end{figure} 

\begin{figure} 
	\caption{An example of the transient dynamics exhibited by mixture A with premixed initial conditions, $\phi=0.50$, showing band splitting, merging and disappearance.} 
	\label{sample-dynamics} 
\end{figure}

\begin{figure} 
	\caption{The exponential growth of the total power in the Fourier spectrum {\it vs.} time measured in rotation periods. The loss of exponentiality corresponds to appearance of saturated bands.  Here $\omega = 3.59$ rad/s.} 
	\label{log-totalpow-with-spacetime} 
\end{figure} 
 
\begin{figure} 
	\caption{The exponential growth rate of the total Fourier power, in units of the inverse rotation period $T=2 \pi/\omega$, {\it vs.} the angular rotation frequency $\omega$ of the tube .} 
	\label{growthrate-vs-freq} 
\end{figure} 

\begin{figure} 
	\caption{The Fourier spectrum at the onset of saturation, averaged over the various $\omega$.} 
	\label{average-fft} 
\end{figure} 

\begin{figure} 
	\caption{ The dominant wavelength at the onset of saturation as a function of $\omega$.  Open symbols show the wavelength at which the largest peak occurs in the Fourier spectrum.  Solid symbols show the centre-of-mass wavelength of the spectrum.} 
	\label{selected-wavelength} 
\end{figure} 

\begin{figure} 
	\caption{Standing waves produced by a presegregated initial condition for $\omega = 4.841$ rad/s and $\phi = 0.67.$  Only a portion of the tube is shown.} 
	\label{seeded} 
\end{figure} 

\begin{figure} 
	\caption{The dispersion relation for traveling waves at tube rotation frequency $\omega = 4.841$ rad/s and $\phi = 0.67.$  The solid symbols indicate presegregated initial conditions. The open symbols are from spontaneous waves using premixed initial conditions, with speeds found from Fourier peaks (open square), and averages of slopes in spacetime images (open circle).} 
	\label{dispersion-relation} 
\end{figure} 

\begin{figure} 
	\caption{Band merging, followed by frozen behavior near the knee in the dispersion relation for traveling waves, for $\omega = 4.841$ rad/s and $\phi = 0.67.$ Only a portion of the tube is shown.} 
	\label{merging} 
\end{figure} 

\begin{figure} 
	\caption{(a) The amplitude of a standing wave {\it vs.} time.  (b) The speed of the left- and right-going components of the traveling wave, {\it vs.} the peak amplitude squared. Here, $\omega = 4.841$ rad/s and $\phi = 0.67.$} 
	\label{vamplitude} 
\end{figure} 

\begin{figure} 
	\caption{The traveling wave speed versus the filled volume fraction $\eta$.  The leftmost point corresponds to $\eta=0.28$, which was used for most measurements.  } 
	\label{speed-vs-filling} 
\end{figure} 

\begin{figure} 
	\caption{The traveling wave speed, expressed in mm per rotation period $T = 2\pi/\omega$ vs. the rotation period $T$.  The waves stop traveling for periods of 1.15 s and lower.} 
	\label{speed-vs-period} 
\end{figure} 

\begin{figure} 
	\caption{``The fountain'': a persistent source of traveling waves which was stable over several thousand rotations.} 
	\label{fountain} 
\end{figure} 

\begin{figure} 
	\caption{Two pulses of traveling waves emerging from a locally presegregated initial condition. Spontaneous traveling waves and stationary bands are also evident. } 
	\label{pulse} 
\end{figure} 

\end{document}